\documentclass[a4paper,11pt]{article}
\usepackage{amssymb}
\usepackage{amsfonts}
\usepackage{amsmath}

\input{tcilatex}

\begin{document}

\title{Determination of total mechanical energy of the universe within the
framework of Newtonian mechanics}
\author{Dimitar Valev \\
%EndAName
\textit{Stara Zagora Department, Solar-Terrestrial Influences Laboratory,}\\
\textit{Bulgarian Academy of Sciences, 6000 Stara Zagora, Bulgaria}}
\maketitle

\begin{abstract}
The recent astronomical observations indicate that the expanding universe
having a finite particle horizon is homogeneous, isotropic and
asymptotically flat. The Euclidean geometry of the universe enables to
determine the total kinetic and gravitational energies of the universe
within the framework of the Newtonian mechanics. It has been shown that
almost the entire kinetic energy of the universe ensues from the
cosmological expansion. Both, the total kinetic and gravitational energies
of the universe have been determined in relation to an observer at arbitrary
location. It is amazing that the modulus of the total gravitational energy
differs from the total kinetic energy with a multiplier close to a unit.
Thus, the total mechanical energy of the universe has been found close to
zero. Both, the total kinetic energy and the modulus of total gravitational
energy of the universe are estimated to $3/10$ of its total rest energy $%
Mc^{2}$.

Key words: energy of the universe; flat universe; Newtonian mechanics
\end{abstract}

\section{Introduction}

\bigskip

The problem for the average density of the universe $\overline{\rho }$
acquires significance when it has been shown that the General Relativity
allows to reveal the geometry and evolution of the universe by simple
cosmological models \cite{Friedman 1922, Lemaitre 1927, Einstein 1932}.
Crucial for the universe appears dimensionless total density $\Omega =\frac{%
\overline{\rho }}{\rho _{c}}$, where $\rho _{c}$\ is the critical density of
the universe. In the case of $\Omega <1$ (open universe) the global spatial
curvature is negative and the geometry of the universe is hyperbolic and in
the case of $\Omega >1$ (closed universe) the curvature is positive and the
geometry is spherical. In the special case of $\Omega =1$ (flat universe)
the curvature is zero and the geometry is Euclidean. Until recently scarce
information has been available about the density and geometry of the
universe. The most trustworthy total matter density $\Omega $ has been
determined by measurements of the dependence of the anisotropy of the Cosmic
Microwave Background (\textit{CMB}) upon the angular scale. The recent
results show that $\Omega \thickapprox 1\mp \Delta \Omega $, where the error 
$\Delta \Omega $ decreases from 0.10 \cite{de Bernardis 2000, Balbi 2000} to
0.02 \cite{Spergel 2003}, i.e. the density of the universe is close to the
critical one and the universe is asymptotically \textit{flat} (Euclidean).
The fact that $\Omega $ is so close to a unit is not accidental since only
at $\Omega =1$\ the geometry of the universe is flat and the flat universe
was predicted from the inflationary theory \cite{Guth1981}. The total
density $\Omega $ includes densities of baryon matter $\Omega
_{b}\thickapprox 0.05$, cold dark matter $\Omega _{c}\thickapprox 0.25$\ 
\cite{Peacock 2001} and dark energy $\Omega _{\Lambda }\thickapprox 0.70$\
producing an accelerating expansion of the universe \cite{Riess 1998,
Perlmutter 1999}.

The found negligible \textit{CMB} anisotropy $\frac{\delta T}{T}\sim 10^{-5}$%
\ indicates that the early universe was very homogeneous and isotropic \cite%
{Bennett 1996}. Three-dimensional maps of the distribution of galaxies
corroborate homogeneous and isotropic universe on large scales greater than $%
100$ $Mps$ \cite{Shectman 1996, Stoughton 2002}.

Usually, Einstein pseudotensor is used for determination of the total energy
of the universe \cite{Rosen 1994, Johri 1995}. This approach is general for
open, close and flat anisotropic models, but pseudotensorial calculations
are dangerous as they are very coordinate dependent and thus, they may lead
to ambiguous results \cite{Banerjee 1997}. Euclidean geometry of the
universe still enables unambiguous determining of the total kinetic and
gravitational energies of the universe, within the framework of classical
mechanics.

\section{Determination of the total gravitational energy of the universe}

The finite time of cosmological expansion (age of the universe) and the
finite speed of light $c$ set a \textit{finite} particle horizon beyond
which no material signals reach the observer. The finite particle horizon
(\textquotedblleft radius\textquotedblright ) of the universe is equal to
Hubble distance $R\sim cH^{-1}$, where $H=H_{0}h\approx 70$ $%
kms^{-1}Mps^{-1} $ \cite{Mould 2000} is the Hubble expansion rate and $%
H^{-1}\thickapprox 14$ $Gyr$ is the Hubble time (age of the universe).
Therefore, for an observer in \textit{arbitrary} location, the universe
appears a three-dimensional, homogeneous and isotropic sphere having finite
radius (particle horizon) $R\sim cH^{-1}$. The possible matter beyond the
particle horizon does not affect the observer.

The gravitational potential energy $U$ of a homogeneous sphere with density $%
\overline{\rho }$\ is:

\begin{equation}
U=-G\tint\limits_{0}^{R}\frac{M(r)dm}{r}=-\frac{16}{3}G\pi ^{2}\overline{%
\rho }^{2}\tint\limits_{0}^{R}r^{4}dr=-\frac{3}{5}\frac{GM^{2}}{R}
\label{Eqn1}
\end{equation}

where $G$ is the gravitational constant, $M(r)=\frac{4}{3}\pi r^{3}\overline{%
\rho }$ is the mass of a sphere with radius $r$ and density $\overline{\rho }
$, $M$ is the mass and $R$ is the radius of the universe.

In consideration of $R\sim cH^{-1}$\ and $\overline{\rho }=\Omega \rho _{c}$%
, we obtain for the mass of the universe:

\begin{equation}
M=\frac{4}{3}\pi R^{3}\overline{\rho }=\frac{4}{3}\frac{\pi c^{3}\Omega \rho
_{c}}{H^{3}}  \label{Eqn2}
\end{equation}

The critical density of the universe \cite{Peebles 1971} is determined from
equation (\ref{Eqn3}):

\begin{equation}
\rho _{c}=\frac{3H^{2}}{8\pi G}  \label{Eqn3}
\end{equation}

Replacing (\ref{Eqn3}) in (\ref{Eqn2}) we have found equation (\ref{Eqn4})
for the mass of the universe including dark matter and dark energy:

\begin{equation}
M=\frac{c^{3}\Omega }{2GH}  \label{Eqn4}
\end{equation}

Finally, replacing (\ref{Eqn4}) in (\ref{Eqn1}) and on account of $R\sim
cH^{-1}$ we have found for the total gravitational energy of the universe:

\begin{equation}
U=-\frac{3}{20}\frac{c^{5}}{GH}\Omega ^{2}  \label{Eqn5}
\end{equation}

Similar approach has been used for calculation of the total gravitational
energy of a body arising from gravitational interaction of the body with all
masses of the observable universe \cite{Woodward 1975, Valev 2009}.

\section{Determination of the total kinetic and total mechanical energies of
the universe}

The estimation of the total kinetic energy of the universe $T$ is
complicated as a result of the diversity of movements of masses in the
universe. We suggest that \textit{almost all} kinetic energy of the universe
is a result of the cosmological expansion since it includes movement of the
enormous masses (galaxies and clusters of galaxies) with average speed of
the order of magnitude of $\frac{c}{2}$. The rotation curves of galaxies
show that the majority of stars move into the galaxies with speed less than $%
v_{0}=3\times 10^{5}$ $m/s$ \cite{Sofue 1996}. Besides, on rare occasions,
the peculiar (non-cosmological) velocities of galaxies exceed this value 
\cite{Strauss 1995}. On the other hand, the speed of medium-distanced
galaxies (and their stars), as a result of the cosmological expansion is of
the order of magnitude of $\frac{c}{2}=1.5\times 10^{8}$ $m/s$. Obviously,
the kinetic energy of an \textquotedblleft average star\textquotedblright\
in the universe, ensuing from its peculiar movement, constitutes less than $(%
\frac{v_{0}}{c/2})^{2}\sim 4\times 10^{-6}$ part of its kinetic energy,
ensuing from the cosmological expansion, therefore, former should be ignored.

Let us estimate the total kinetic energy of the universe in relation to an
observer at \textit{arbitrary} location. The total kinetic energy of the
universe is the sum of the kinetic energy of all masses $m_{i}$ moving in
relation to the observer with speed $v_{i}$\ determined from Hubble law $%
v_{i}=Hr_{i}$, where $r_{i}\leq cH^{-1}$\ is the distance between the
observer and mass $m_{i}$\ placed within the particle horizon:

\begin{equation}
T=\frac{1}{2}\tsum\limits_{i}m_{i}v_{i}^{2}  \label{Eqn6}
\end{equation}

Although distant galaxies escape from the observer with speeds $v_{i}$
comparable to $c$, the Newtonian formula (\ref{Eqn6}) of kinetic energy is
used. Thus, all calculations are made within the framework of the Newtonian
mechanics.

Since for an arbitrary observer the universe appears a three-dimensional,
homogeneous and isotropic sphere having finite radius (particle horizon) $%
R\sim cH^{-1}$, the sum (\ref{Eqn6}) can be replaced by integral (\ref{Eqn7}%
):

\begin{equation}
T=\frac{1}{2}\tint mv^{2}dr=\frac{1}{2}\tint\limits_{0}^{R}4\pi r^{2}\rho
v_{r}^{2}dr=2\pi \overline{\rho }\tint\limits_{0}^{R}v_{r}^{2}r^{2}dr
\label{Eqn7}
\end{equation}

Replacing $v_{r}$\ with expression from Hubble law $v_{r}=Hr$, equation (\ref%
{Eqn7}) is transformed into (\ref{Eqn8}):

\begin{equation}
T=2\pi \overline{\rho }H^{2}\tint\limits_{0}^{R}r^{4}dr=\frac{2}{5}\pi 
\overline{\rho }H^{2}R^{5}  \label{Eqn8}
\end{equation}

In consideration of $R\sim cH^{-1}$ we obtain:

\begin{equation}
T=\frac{2}{5}\frac{\pi \overline{\rho }c^{5}}{H^{3}}  \label{Eqn9}
\end{equation}

Since $\overline{\rho }=\Omega \rho _{c}$ (\ref{Eqn9}) is transformed into (%
\ref{Eqn10}):

\begin{equation}
T=\frac{2}{5}\frac{\pi \Omega \rho _{c}c^{5}}{H^{3}}  \label{Eqn10}
\end{equation}

After replacing (\ref{Eqn3}) in (\ref{Eqn10}) we have found the equation for
the total kinetic energy of the universe:

\begin{equation}
T=\frac{3}{20}\frac{c^{5}}{GH}\Omega  \label{Eqn11}
\end{equation}

It is amazing that the equations for the total gravitational and kinetic
energies of the universe (\ref{Eqn5}) and (\ref{Eqn11}) differ one from
another by a multiplier $\Omega \thickapprox 1$.

Clearly, the total mechanical energy of the universe $E$ is:

\begin{equation}
E=T+U=\frac{3}{20}\frac{c^{5}}{GH}\Omega (1-\Omega )  \label{Eqn12}
\end{equation}

Since the resent \textit{CMB} measurements obtain $\Omega \thickapprox 1$,
then $E=T+U\thickapprox 0$, i.e. the total mechanical energy of the universe
is close to \textit{zero}. This substantial result confirms the conjecture
that the gravitational energy of the universe is approximately balanced with
its kinetic energy of the expansion \cite{Lightman 1984}.

Formulae (\ref{Eqn5}) and (\ref{Eqn11}) show that the total kinetic energy
of the universe is close to the modulus of its total gravitational energy:

\begin{equation}
T\thickapprox \left\vert U\right\vert \thickapprox \frac{3}{20}\frac{c^{5}}{%
GH}  \label{Eqn13}
\end{equation}

From Einstein equation and (\ref{Eqn4}) we can determine the total rest
energy of the universe:

\begin{equation}
E_{0}=Mc^{2}=\frac{1}{2}\frac{c^{5}}{GH}\Omega \thickapprox \frac{1}{2}\frac{%
c^{5}}{GH}  \label{Eqn14}
\end{equation}

Therefore, the total kinetic energy of the universe is $\frac{3}{10}$ of its
total rest energy $E_{0}$. The same is valid for the modulus of the total
gravitational energy of the universe, too.

\section{Discussions}

Formulae (\ref{Eqn13}) and (\ref{Eqn14}) show that the total gravitational
energy of the universe is $U=-\frac{3}{10}E_{0}=-\frac{3}{10}Mc^{2}$, i. e.
30 \% of the total rest energy of the universe. On the other hand, the
gravitational energy of a sphere having mass $m$ and radius $r$ equal to its
Schwarzschild's radius $r_{s}=\frac{2Gm}{c^{2}}$\ is:

\begin{equation}
U=-\frac{3}{5}\frac{Gm^{2}}{r_{s}}=-\frac{3}{10}mc^{2}  \label{Eqn15}
\end{equation}

which also appears 30 \% of the total rest energy of the mass $m$.

This suggests that the particle horizon (\textquotedblleft
radius\textquotedblright ) of the universe ($R$) coincides with
Schwarzschild's radius of the universe ($R_{s}$). It is easy to prove this
statement using the following:

On account of $\Omega =\frac{\overline{\rho }}{\rho _{c}}\thickapprox 1$ and 
$H\sim cR^{-1}$ we obtain:

\begin{equation}
\Omega =\frac{\overline{\rho }}{\rho _{c}}=\frac{3M/4\pi R^{3}}{3H^{2}/8\pi G%
}=\frac{2GM}{R^{3}H^{2}}\sim \frac{2GM}{Rc^{2}}\thickapprox 1  \label{Eqn16}
\end{equation}

Therefore, $R\sim cH^{-1}\thickapprox \frac{2GM}{c^{2}}\equiv R_{s}$, i. e.
the particle horizon (\textquotedblleft radius\textquotedblright ) of the
universe \textit{coincides} with Schwarzschild's radius of the universe.

It is interesting that formulae for the total mass and rest energy of the
universe (\ref{Eqn4}) and (\ref{Eqn14}) would be found by dimensional
analysis, without consideration of the total density of the universe. In
fact, a quantity with mass dimension $M=kc^{\alpha }G^{\beta }H^{\gamma }$\
could be composed by using the fundamental parameters - speed of the light ($%
c$), gravitational constant ($G$) and Hubble constant ($H$). Dimensional
analysis gave $\alpha =3$ and $\beta =\gamma =-1$ . Therefore:

\begin{equation}
M=k\frac{c^{3}}{GH}  \label{Eqn17}
\end{equation}

where $k\sim 1$.

Analogously, a quantity with dimension of energy $E_{0}=k_{0}c^{a}G^{b}H^{d}$%
\ could be composed by parameters $c$, $G$ and $H$. In this case dimensional
analysis has shown that $a=5$ and $b=d=-1$. Therefore,

\begin{equation}
E_{0}=k_{0}\frac{c^{5}}{GH}  \label{Eqn18}
\end{equation}

where $k_{0}\sim 1$.

Obviously, the formulae (\ref{Eqn17}) and (\ref{Eqn18}) coincide with
formulae of total mass and rest energy of the universe (\ref{Eqn4}) and (\ref%
{Eqn14}), with accuracy to a factor $k=\frac{1}{2}$. Thus, the total mass
and rest energy of the universe have been found by original independent
approach, namely dimensional analysis, which reinforces our presumptions for
accepted cosmological model.

\section{Conclusions}

The recent astronomical observations indicate that the expanding universe
having a finite particle horizon is homogeneous, isotropic and flat. The
Euclidean geometry of the universe enables to determine the total kinetic
and gravitational energies of the universe within the framework of the
Newtonian mechanics.

Based on the simple homogeneous and isotropic model of the flat universe
which expands according to Hubble law, we have found equations (\ref{Eqn5})
and (\ref{Eqn11}) for the total gravitational and kinetic energy of the
universe, respectively. It is amazing that these equations differ by a
multiplier $\Omega \thickapprox 1$. That is, the modulus of the total
gravitational energy of the universe is close to its total kinetic energy
and therefore the total mechanical energy $E$ of the universe is close to
zero, i. e. the gravitational energy of the universe is balanced with its
kinetic energy of the expansion.

Both, the total kinetic and gravitational energies of the universe have been
determined in relation to an observer at arbitrary location. The total
kinetic energy and the modulus of the total gravitational energy of the
universe are estimated to $\frac{3}{10}$ of its total rest energy $Mc^{2}$.

Finally, it has been shown that the particle horizon (\textquotedblleft
radius\textquotedblright ) of the universe coincides with Schwarzschild's
radius of the universe. Besides, the total mass and rest energy of the
universe have been estimated by dimensional analysis, without consideration
of total density of the universe.

\bigskip

\end{document}